\documentclass[12pt]{iopart}

\usepackage{graphicx}
\usepackage{bm}
\usepackage{amssymb}     
\usepackage{epsfig}
\usepackage{pstricks,pst-grad}
\usepackage{bbm}

\begin{document}

\title{Stable, metastable and unstable states in the mean-field RFIM at $T=0$}

\author{M.L.~Rosinberg}
%
\address{Laboratoire de Physique Th\'eorique de la Mati\`ere Condens\'ee, Universit\'e Pierre et Marie Curie\\ 4 place Jussieu, 75252 Paris Cedex 05, France}

\author{G.~Tarjus}
\address{Laboratoire de Physique Th\'eorique de la Mati\`ere Condens\'ee, Universit\'e Pierre et Marie Curie\\ 4 place Jussieu, 75252 Paris Cedex 05, France}

\author{F.J.~P\'erez-Reche}
%
\address{Department of Chemistry, University of Cambridge, Cambridge, CB2 1EW, UK}

\begin{abstract}
We compute the probability  of finding metastable states at a given field in the mean-field random field Ising model  at $T=0$. Remarkably, this probability is finite in the thermodynamic limit, even on the so-called ``unstable'' branch of the magnetization curve. This implies that the branch is reachable when the magnetization is controlled instead of the magnetic field, in contrast with the situation in the pure system.
\end{abstract}

\maketitle

\def\be{\begin{equation}}
\def\ee{\end{equation}}
\def\bea{\begin{eqnarray}}
\def\eea{\end{eqnarray}}

\section{Introduction}

As is well known, the main approximation underlying  mean-field theories of  phase transitions consists in neglecting any spatial dependence of the order parameter. In simple systems such as fluids or ferromagnets this leads to an equation of state (e.g. the van der Waals equation) that exhibits a continuous loop below the critical temperature $T_c$. In this framework, one can distinguish between  stable (equilibrium), metastable, and unstable states. The intermediate, unstable branch of the loop is associated with maxima of the free energy and has a negative slope  (whence a negative susceptibility), and the unstable and metastable regions of the phase diagram are separated by a spinodal line where the free-energy barrier vanishes. Whereas the concept of metastability can be in some sense extended to systems with short-range interaction in finite dimensions where it becomes a matter of timescales, the unstable branch is a complete artifact of the mean-field approximation and the {\it homogeneous}  states along this branch do not represent any real physical situation. 

Things are different in the presence of disorder since inhomogeneities are induced even at a microscopic scale. As a result, configurations formed by a multitude of small domains of the two  phases
can become metastable in an extended range of the ``external'' field (e.g.  the fluid pressure or the magnetic field). Interestingly, this can already occur at the mean-field level and the purpose of this paper is to emphazise  that metastable states are present {\it all along }  the so-called unstable branch in the mean-field random field Ising model (RFIM) at $T=0$. This feature is in fact a precursor of what happens in the  RFIM in finite dimensions, and it has  consequences for experiments in actual random-field systems at  low temperature.

\section{The $T=0$ RFIM on a fully-connected lattice}

We consider a collection of $N$ Ising spins ($s_i=\pm 1$) interacting   via the Hamiltonian
\be
\label{Eq01} 
{\cal H}=-\frac{J}{2N}\sum_{i\ne j}s_i s_j-\sum_i (h_i+H)s_i
\ee
where $J>0$, $H$ is a uniform external field,  and $\{h_i\}$ is a collection of random fields drawn identically and independently from some probability distribution ${\cal P}(h)$. (In the following, we consider a Gaussian distribution with zero mean and standard deviation $\Delta$.)  This mean-field model can be obtained from the usual short-range RFIM by placing the spins on a fully-connected lattice (the connectivity is then equal to $N-1$ and the exchange interaction is rescaled by $N$ to ensure a proper thermodynamic limit). Eq. (\ref{Eq01}) can be also rewritten as
\be
\label{Eq02} 
{\cal H}=-\sum_i f_i s_i
\ee
where $f_i=J(m-s_i/N)+h_i+H$ is the effective field acting on each spin $i$ and $m=(\sum_i s_i)/N$ is the magnetization per spin. 

It is straightforward to compute the equilibrium properties of this model by the replica method, without all the complications that plague the case of random exchange. In particular, the average magnetization  in the thermodynamic limit $N \rightarrow \infty$ is solution of the self-consistent equation\cite{SP1977}

\be
\label{Eq03} 
m(H)=\int {\cal P}(h) \tanh[\beta (Jm(H)+H+h)]dh
\ee
where $\beta=1/k_BT$. (In Ref.\cite{SP1977}, only the case $H=0$ is considered, but the generalization to $H\ne 0$ is straightforward.) At $T=0$, the equation becomes

\bea
\label{Eq04} m_0(H)&=\int {\cal P}(h)\  \mbox{sgn}(Jm_0(H)+H+h)dh\nonumber\\
&= 2p(m_0(H))-1
\eea
where $\mbox{sgn}(x)=x/\vert x\vert$ and $p(m)=\int_{-H-Jm}^{\infty}{\cal P}(h)dh$. This yields  $m_0(H)=\mbox{erf}([H+Jm_0(H)]/\Delta \sqrt{2})$ in the case of the Gaussian distribution, where $\mbox{erf}(x)$ is the error function. Since the magnetization per spin is a self-averaging quantity, $m_0(H)$ is also the expected value of the magnetization in a very large sample.

\begin{figure}[ht]
\begin{center}
\epsfig{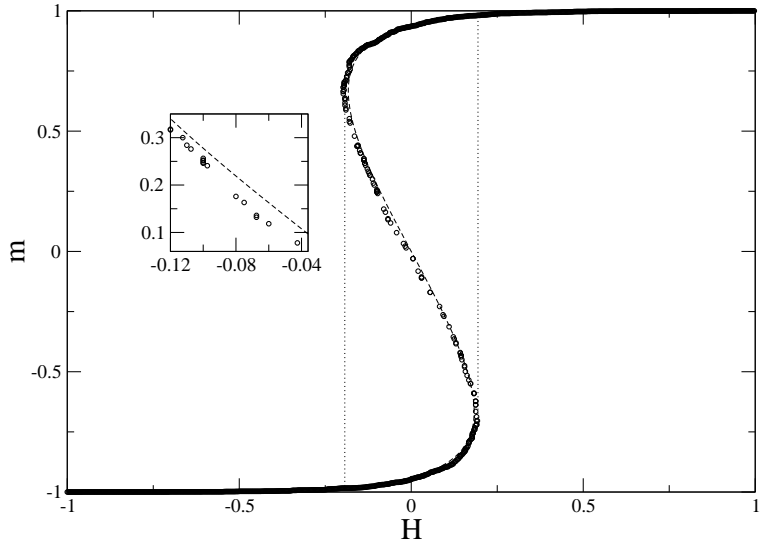}
\end{center}
\caption{\label{Fig 1} Metastable states in the mean-field Gaussian RFIM for a single disorder realization of size $N=5000$ with $\Delta=0.5$ (the increment in the field is $\Delta H=25. 10^{-4}$). The dashed curve represents the solution of Eq. (\ref{Eq04}). The inset shows that there may exist a few states with different magnetizations at the same field.}
\end{figure}\normalsize

Below the critical disorder $\Delta_c^0$ ($\Delta_c^0=\sqrt{2/\pi}J$ for the Gaussian distribution), Eq. (\ref{Eq04}) has three solutions in a certain range of the field and the curve $m_0(H)$ exhibits a characteristic ``van der Waals'' loop with an intermediate branch along which $\partial m/\partial H<0$ (see Fig. 1; we take $J=1$ in all figures). As usual, this behavior is associated with the non-convexity of the  (free) energy. At a given field $H$, the ground-state corresponds to the solution with the lowest overall energy whereas the intermediate branch has to the largest energy. In the space of replica magnetizations\cite{SP1977}, such branch corresponds to an absolute  maximum. The result from the replica method can be shown to be exact\cite{SP1977}, and it seems natural to describe the intermediate branch as ``unstable'',  naively transposing the situation found in the pure system (i.e. in the absence of random field). However, this is misleading. Indeed, since Eq. (\ref{Eq04}) results from an average over disorder (which restores translational invariance) when $N\rightarrow\infty$, it says nothing about the orientation of the spins in a given (finite $N$) sample. It turns out that some of the states along the intermediate branch are metastable (i.e. {\it local} minima of the energy) and not unstable. Indeed,  Eq. (\ref{Eq04}) is trivially verified if 
\be
\label{Eq05} 
s_i=\mbox{sgn}(f_i)_ , \  i=1...N ,
\ee
which is  the definition of  the so-called one-spin-flip stable states. There is nothing  new here: it is  known that  Eq. (\ref{Eq04}) also describes the nonequilibrium (hysteretic) behavior of the mean-field RFIM at $T=0$ with the single-spin-flip (Glauber) dynamics\cite{S1993}. As $H$ is slowly varied from $\pm \infty$, this dynamics imposes at any field that the spins with $h_i<-Jm-H$ point down whereas the other ones point up, and the self-consistent equation for the average magnetization, $m(H)=\int {\cal P}(h)s_i dh$,  is just Eq.(\ref{Eq04}). (Note incidentally that this corresponds to an ``annealed'' average, but it yields the correct result in this case.)  The nonequilibrium   RFIM  at $T=0$  has been the subject of extensive studies in recent years\cite{SD} and the mean-field model, despite some peculiar features (the equilibrium and nonequilibrium critical disorders coincide and there is no hysteresis for $\Delta>\Delta_c^0$), has the advantage that many interesting properties can be computed exactly (for instance the size and duration of avalanches)\cite{S1993,SJK2006}. For $\Delta<\Delta_c^0$, the nonequilibrium system explores the lower (or upper) branch of the loop until $H$  reaches one of the two coercive fields where $dm/dH$ diverges and the magnetization jumps discontinuously. This ``infinite avalanche'' \cite{S1993} occurs when each flipping spin triggers one other spin on average, which corresponds to  the condition $2J{\cal P}(-Jm_0(H)-H)=1$. Hereafter, we shall use the shorthand  ${\cal P}^*_0$ for the  quantity ${\cal P}(-Jm_0(H)-H)$.

The point that we want to emphasize here is that a configuration where each spin satisfies Eq. (\ref{Eq05}) is  a metastable state by definition, {\it whatever} the value taken by $m_0(H)$ (of course, a  state on the intermediate branch has a larger energy than the corresponding ground state and it cannot be reached by controlling the field). This is illustrated in Fig. 1 that shows the magnetizations of the metastable states in a single sample of size $N=5000$ with $\Delta=0.5$ as a function of $H$. One can see that there are  stable states in the intermediate region and that they gather along a curve that will become the  so-called ``unstable''  branch in the thermodynamic limit. At a given field $H$, the number of these  states is very small, obviously not exponentially growing with system size. However, the probability of finding metastable states remains finite when $N \rightarrow \infty$, as we now show.

\section{Metastable states}

For a disorder realization of size $N$, consider the ensemble of metastable configurations  at the field $H$ with exactly $P$ spins up (and thus an overall magnetization $M=2P-N$). The number of such configurations is
\begin{eqnarray}
\label{Eq4}
{\cal N}(M,H)=\mbox{Tr}_{\{s_i\}}  \prod_i \Theta(s_i f_i) \delta_K(\sum_i s_i-M)\  ,
\end{eqnarray}
where $\Theta(x)$ is the Heaviside step function and $\delta_K$ is the Kronecker-$\delta$.  It is easy to see that there cannot be more than one metastable state with magnetization $M$ in a single realization (i.e. ${\cal N}(M,H)=0$ or $1$). Averaging over disorder, we have
\be
\label{EqA1}
\overline{{\cal N}(M,H)}=\big(_P^N\big)p(m-1/N)^P\big[1-p(m+1/N)\big]^{N-P}\ ,
\ee
where the two terms in the right-hand side represent the probabilities of having $P$ spins up and $N-P$ spins down at the field $H$, respectively. Using the Stirling approximation for the factorial and expanding $p(m\pm 1/N)$ to first order in $1/N$, we find
\be
\label{EqA2}
\overline{{\cal N}(M,H)}\sim \sqrt{\frac{2}{\pi N}}\frac{e^{N\phi(m)}}{\sqrt{1-m^2}}e^{-J{\cal P}(-H-Jm)[\frac{1+m}{2p(m)}+\frac{1-m}{2[1-p(m)]}]}
\ee
with
\be
\label{EqA3}
\phi(m)=\frac{1+m}{2}\ln \frac{2p(m)}{1+m}+\frac{1-m}{2}\ln \frac{2[1-p(m)]}{1-m} \ .
\ee
This number is exponentially small  ($\phi(m)<0$) except when $m=2p(m)-1\equiv m_0(H)$, i.e. when $m$ is  solution of Eq.(\ref{Eq04}): $\phi(m)$ is then maximum and equal to $0$.   Expanding $\phi(m)$ to second order close to $m=m_0(H)$, we  find that the average number of metastable states at the field $H$ is finite when $N \rightarrow \infty$ and is given by
\begin{eqnarray}
\label{EqA4}
\overline{{\cal N}(H)}&=\sum_M \overline{{\cal N}(M,H)}\sim\frac{N}{2}\int dm \ \overline{{\cal N}(M,H)}\nonumber\\
&\rightarrow\frac{e^{-2J{\cal P}^*_0}}{\vert 1-2J{\cal P}^*_0\vert}  \ .
\end{eqnarray}

This result is valid both above and below $\Delta_c^0$, and for $\Delta<\Delta_c^0$ the three branches of $m_0(H)$ must be considered separately (with $2J{\cal P}^*_0\ge 1$ along the intermediate branch).  The behavior of $\overline{{\cal N}(H)}$ as a function of $m_0(H)$ is shown in Fig. 2 for $\Delta/J=0.5$. It is worth noting that $\overline{{\cal N}(H)}$ diverges at the coercive fields (i.e. at the spinodal endpoints) where 
  $2p'(m)=2J{\cal P}^*_0=1$ and  $\phi''(m_0)=-[1-2J{\cal P}^*_0]^2/(1-m_0^2)=0$ (from now on, the dependence of $m_0$ on $H$ will not be indicated for brevity). In this case, the large-deviation function $\phi(m)$ must be expanded to  fourth order about $m=m_0$ and $\overline{{\cal N}(H)}$  scales like $N^{1/4}$.  At the critical point ($\Delta=\Delta_c^0$ and $ H=m_0=0$), $\phi(m)$ must be expanded to the sixth order and $\overline{{\cal N}(H)}$  scales like $N^{1/3}$. The special form of $\phi(m)$ is responsible for these unusual mean-field exponents.

\begin{figure}
\begin{center}
\epsfig{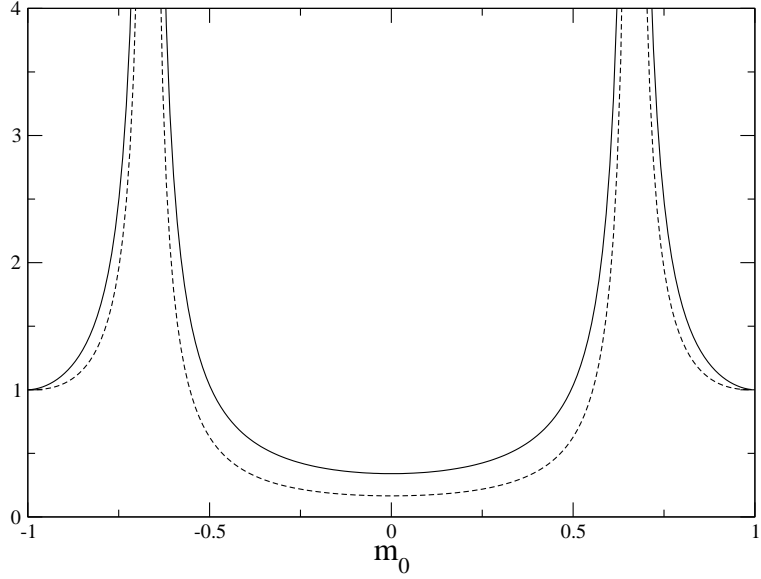}
\end{center}
\caption{\label{Fig 3} Average number of $1$-spin-flip (solid line) and $2$-spin-flip (dashed line) stable states along  the curve $m_0(H)$ for $\Delta=0.5$. Both quantities diverge at the spinodal endpoints.}
\end{figure}

Note that it is crucial to take into account the contributions of order $1/N$ in the local fields $f_i$  in order to  obtain the correct result. This is due to the fact that the  function  $\phi(m)$ is zero at the saddle-point,  so that the associated complexity is zero. (This is different from the situation found in the mean-field spin glass model discussed in Ref.\cite{BM2000}.) On the other hand, the subdominant terms play no role in determining  the probability of finding  a metastable state with magnetization $m$ at the field $H$, which is only given as usual by the fluctuations around the saddle point:
\be
\label{EqA5}
\frac{\overline{{\cal N}(m,H)}}{\overline{{\cal N}(H)}}\sim
\sqrt{\frac{-N\phi''(m_0)}{2\pi}} e^{\frac{N}{2}\phi''(m_0)(m-m_0)^2} \ .
\ee
 This becomes a $\delta$-distribution when $N\rightarrow\infty$ if $\phi''(m_0)\ne 0$.

More generally, we  can  compute all the moments $\overline{{\cal N}(H)^n}$ and use this information to obtain the full probability distribution $P(q)=\overline{\delta_K({\cal N}(H)-q)}$ (with $\overline{{\cal N}(H)^n}\equiv \overline{q^n}=\sum q^n P(q)$).  This amounts to  count all possible ways of ordering $n$ magnetizations $M_1,M_2...M_n$, given that ${\cal N}(M_i,H){\cal N}(M_j,H)={\cal N}(M_i,H)$ when $M_i=M_j$ since ${\cal N}(M,H)$ is just $0$ or $1$. This yields
\bea
\label{EqA7}
\fl\overline{{\cal N}(H)^n}&=\sum_{r=1}^n \sum_{n_1,n_2,...n_r\ge 1}(n_1,n_2...n_r)!\sum_{M_1,M_2...,M_r}\nolimits^> \overline{{\cal N}(M_1,H){\cal N}(M_2,H)...{\cal N}(M_r,H)}
\eea
where $(n_1,n_2...n_r)!=n!/(n_1!n_2!...n_r!)$ is a multinomial coefficient and the sum runs over all $(n_1,n_2,...n_r)$ such that $\sum_{i=1}^r n_i=n$ and $n_1,n_2...n_r\ge 1$.  The notation $\sum\nolimits^>$  indicates that the sum over the magnetizations  $M_1,M_2...,M_r$ is restricted to a specific order,  say $M_r>M_{r-1}>...M_2>M_1$.  We then define $M_i=2P_i-N$, where $P_i$ is the number of spins up, and introduce the (strictly)  positive quantities $Q_i=P_{i}-P_{i-1}$ ($i=2,...r$). It is  easy to see that Eq.  (\ref{EqA1})   generalizes to 
\bea
\label{EqA8}
\overline{{\cal N}(M_1,H){\cal N}(M_2,H)...{\cal N}(M_r,H)}=N!\frac{ p(m_1-1/N)^{P_1}}{P_1!}\nonumber\\
\fl\times\prod_{i=2}^r\frac{[p(m_i-1/N)-p(m_{i-1}+1/N)]^{Q_i}}{Q_i !}
\frac{[1-p(m_r+1/N)]^{N-P_1-\sum_{i=2}^r Q_i}}{(N-P_1-\sum_{i=2}^rQ_i)!} \ .
\eea
Since only the magnetizations in the close neighborhood of $m_0$ contributes to the sum $\Sigma ^>$ when $N \gg 1$ (with the $Q_i$'s  being at most of the order  $N^{\alpha}$ with $0<\alpha<1$),  we can expand  $p(m_i\pm 1/N)$ around $p(m_0)$ to first order in $1/N$ and use $1-P_1/N\sim (1-m_0)/2=1-p(m_0)$.  After some straightforward manipulations, we then obtain 
\bea
\label{EqA13}
\lim_{N \rightarrow \infty}\sum_{M_1,M_2...,M_r}  \nolimits^> \overline{{\cal N}(M_1,H){\cal N}(M_2,H)...{\cal N}(M_r,H)}=\overline{{\cal N}(H)}\nonumber\\
\times\Big[\sum_{Q\ge 1} \frac{(Q-1)^Q}{Q!} (2J{\cal P}^*_0e^{-2J{\cal P}^*_0})^Q\Big]^{r-1} 
\eea
and, after inserting this expression into Eq. (\ref{EqA7}), 
\be
\label{EqA14}
\frac{\overline{{\cal N}(H)^n}}{\overline{{\cal N}(H)}}=\sum_{r=1}^n \sum_{n_1,n_2,...n_r\ge 1}(n_1,n_2,...n_r)![a(m_0)-1]^{r-1}
\ee
where  
\be
\label{EqA15}
a(m_0)=\sum_{k\ge 0}\frac{(k-1)^k}{k!}(2J{\cal P}^*_0e^{-2J{\cal P}^*_0})^k \ . 
\ee
We recognize in (\ref{EqA15}) the series expansion of  $z/[W(z)(1+W(z))] $ near the origin, where  $W(z)$ is the so-called Lambert function, defined as the root of the equation $W(z)e^{W(z)}=z$\cite{CGHJK1996}.  This series converges for $\vert z\vert<1/e$, which is always true in Eq. (\ref{EqA15}) where $z=-2J{\cal P}^*_0e^{-2J{\cal P}^*_0}$.  (The series (\ref{EqA15}) only refers to the principal branch $W_0(z)$ which takes on values between $-1$ to $+\infty$ for $z\ge -1/e$ and is analytic at $z=0$.) As a result, 
\begin{eqnarray}
\label{EqA16}
a(m_0)=\frac{-2J{\cal P}^*_0e^{-2J{\cal P}^*_0}}{W_0(-2J{\cal P}^*_0e^{-2J{\cal P}^*_0})[1+W_0(-2J{\cal P}^*_0e^{-2J{\cal P}^*_0})]} \ ,
\end{eqnarray} 
which yields
\begin{eqnarray}
\label{EqA17}
a(m_0)&=\frac{e^{-2J{\cal P}^*_0}}{1-2J{\cal P}^*_0}\equiv  \overline{{\cal N}(H)} \ \  \mbox{if $2J{\cal P}^*_0<1$} \, ,\nonumber\\
a(m_0)&=\frac{e^{W_0(-2J{\cal P}^*_0e^{-2J{\cal P}^*_0})}}{1+W_0(-2J{\cal P}^*_0e^{-2J{\cal P}^*_0})}\ \  \mbox{if $2J{\cal P}^*_0> 1$}  \ .
\end{eqnarray} 

Knowing from Eq. (\ref{EqA14}) all the moments $\overline{q^n} $ of the probability distribution $P(q)$, we can  build the generating function
\be
\label{EqA18}
\sum_{q\ge 0}e^{-\lambda q}P(q)=1+\sum_{n\ge 1} (-\lambda)^n{\overline q^n} 
\ee
to obtain
\be
\label{EqA19}
\sum_{q\ge 0}e^{-\lambda q}P(q)=\frac{1+(a-{\overline q})(e^{\lambda}-1)}{1+a(e^{\lambda}-1)}
\ee
with ${\overline q} \equiv \overline{{\cal N}(H)}$. This equation can be inverted, showing that $P(q)$ decreases exponentially  for $q\ge 1$. More precisely, we have
\begin{eqnarray}
\label{EqA20}
P(0)&=1-\frac{{\overline q}}{a}\ , \nonumber\\
P(q)&=\frac{{\overline q}}{a(a-1)}\big(\frac{a-1}{a}\big)^q\ \  \mbox{for $q\ge 1$} \ .
\end{eqnarray} 
where both ${\overline q}$ and $a$ are functions of $m_0(H)$. 

For $\Delta>\Delta_c^0$,  $a={\overline q}$ so that  $P(0)=0$ and the most probable value of ${\cal N}(H)$ is  $q=1$, as could be expected. (What is perhaps less expected\cite{note0} is that $P(q)\ne 0$ for $q>1$ and that ${\overline q}=\overline{{\cal N}(H)}>1$.)
For  $\Delta<\Delta_c^0$,  the most probable value  of ${\cal N}(H)$ on the intermediate branch is $q=0$ except very close to the  spinodal end-points where it is again $q=1$ (for $1<2J{\cal P}^*_0< 1.0073$, $P(1)={\overline q}/a^2>P(0)$). Note also that $P(q)$ decreases more and more slowly  when approaching the spinodals as the  inverse characteristic scale $\xi^{-1}=\ln(1-1/a)\rightarrow 0$.
In all cases,  there is  a finite probability of finding a few metastable states at a given field\cite{note1}, as illustrated in Fig. 3 that results from an exact enumeration of  all metastable states in $5000$ disorder realizations of size $N=20000$ at $H=0$ for $\Delta=1$ and $\Delta=0.5$ (in the latter case, only the states in the vicinity of the intermediate branch, i.e. around $m_0=0$, are counted). One has $a={\overline q}\approx 2.12$ for $\Delta=1$, and ${\overline q}\approx 0.340$, $a\approx 1.324$  for $\Delta=0.5$. The numerical data shown in the Figure are in very good agreement with the predictions of Eq. (\ref{EqA20}). 

\begin{figure}
\begin{center}
 \epsfig{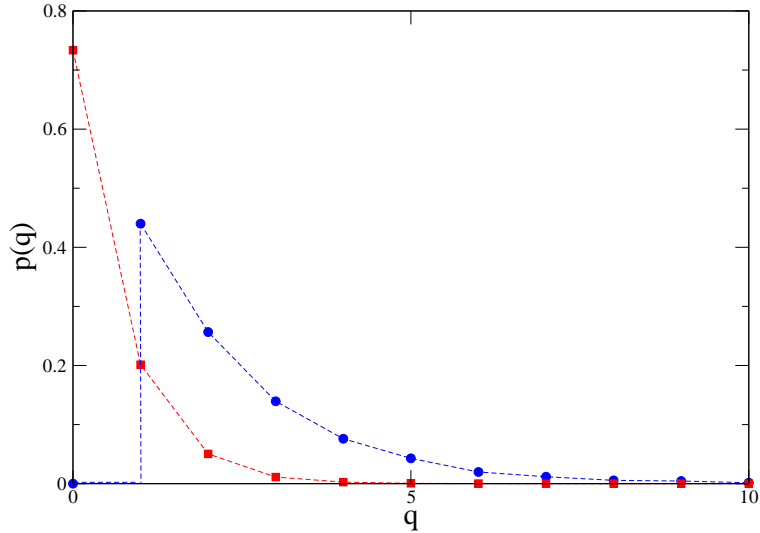}
\end{center}
\caption{\label{Fig 2} Probability  of finding $q$ metastable states in zero field for $\Delta=1$ (blue circles) and $\Delta=0.5$ (red squares) in the mean-field RFIM (the statistics is taken over $5000$ disorder realizations of size $N=20000$). For $\Delta=0.5$, only the states in the vicinity of the intermediate branch (around $m=0$) are counted. The dashed lines are guides for the eye. (Color on line)}
\end{figure}

The above calculations can be generalized to  $2, 3,...k$-spin-flip stable states, i.e. to spin configurations whose energy cannot be lowered by the flip of any subset of $1,2,...k$ spins\cite{NS1999}.  It is easy to see  that a configuration with $P$ spins up and $N-P$ spins down is $k$-stable if it is $(k-1)$-stable and if  the random fields on the $P$ spins up satisfy  $\sum_{\alpha=1}^kh_{i_{\alpha}}>-k(Jm+H)+k^2J/N$ whereas the fields on the $N-P$ spins down satisfy $\sum_{\alpha=1}^k h_{i_{\alpha}}<-k(Jm+H)-k^2J/N$ for any subset $\{i_1,i_2,...i_k\}$. From this, one can  for instance compute the  average number of  $2$-spin-flip stable states  at the field $H$ and find that
\begin{eqnarray}
\label{EqA21}
\overline{{\cal N}^{(2)}(H)}\rightarrow\frac{[2e^{-2J{\cal P}^*_0}-e^{-3J{\cal P}^*_0}]^2}{\vert 1-2J{\cal P}^*_0\vert}  
\end{eqnarray}
when $N\rightarrow \infty$.  The comparison with $\overline{{\cal N}^{(1)}(H)}\equiv\overline{{\cal N}(H)}$  is shown in Fig. 2.  Of course, one has $\overline{{\cal N}^{(k)}(H)}\le\overline{{\cal N}^{(k-1)}(H)}...\le\overline{{\cal N}^{(1)}(H)}$. We have not investigated the behavior for $k\sim\sqrt N$\cite{note2}.
In any case, in order to go from a metastable state on the intermediate branch to the ground state on the lower or upper branch, one needs to flip a number of spins of order $N$ (which corresponds to take into account another solution of the saddle-point equation).

\section{Conclusion}

In this paper, we  have computed the probability of finding metastable states along the so-called  ``unstable''  branch of the 
mean-field RFIM at $T=0$ and shown that the average number remains finite in the thermodynamic limit. 

The presence of a few metastable states  along the intermediate part of the curve $m_0(H)$ for  $\Delta<\Delta_c$ can be considered as a precursor of the phenomenology observed in the $T=0$ RFIM with finite-range  exchange interaction.  This will be investigated in a forthcoming paper\cite{PRT2008b} dealing with random graphs of large but finite connectivity $z$.   Preliminary results  indicate that, as soon as $z$ is finite, a strip of finite width develops around the curve $m_0(H)$ in the field-magnetization plane,  strip  in which the density of the typical metastable states scales exponentially with the system size. This occurs both above and below $\Delta_c$.  As the connectivity decreases, one expects the strip to widen but to remain distinct from the actual hysteresis loop in the weak-disorder regime. In fact, as suggested  in Refs.\cite{DRT2005,PRT2008a},
one may associate the discontinuity in the hysteresis loop below $\Delta_c$ to the existence of a  gap in the magnetization of the metastable states beyond a certain value of the field.  Of course, all this is strictly valid only at $T=0$. However, as is well known, free-energy barriers are very large in random-field systems and thermally activated processes  are not expexted to play a significant role on experimental time scales, at least at low  temperature. Therefore, the above picture is expected to be relevant to real situations. In particular, the  presence of metastable states (and not simply unstable ones as in pure systems) in the central part of the hysteresis loop means that this region could be experimentally accessible, for instance by controlling the magnetization instead of the magnetic field (and more generally the  extensive variable conjugated to the external field). This can be put in relation with the re-entrant hysteresis loops that are observed in some magnetic systems\cite{B1998} or in shape-memory alloys\cite{BMPREV2007} (see also the discussion in Ref.\cite{IRV2006}).


\section{References}

\begin{thebibliography}{10}
\bibitem{SP1977} Schneider T and Pytte E,  1977 {\it Phys. Rev. B. }{\bf 15}, 1519
\bibitem{S1993} Sethna J P, Dahmen K A, Kartha S, Krumhansl J A,
Roberts B W and Shore J D, 1993  {\it Phys. Rev. Lett.} {\bf 70}, 3347; Dahmen K and  Sethna J P, 1996 {\it Phys. Rev. B} {\bf 53}, 14872
\bibitem{SD} Sethna J P, Dahmen K A and Perkovi\'c 0, 2006 in {\it The Science of Hysteresis II}, edited by Bertotti G and Mayergoyz I, Academic Press, Amsterdam
\bibitem{SJK2006} Spasojevi\'c  Dj, Jani\'cevi\'c  S and Kne\~zevi\'c M, 2006  {\it Europhys. Lett.} {\bf 76}, 912
\bibitem{BM2000}  Biroli G and Monasson R,  2000 {\it Europhys. Lett.} {\bf 50}, 155
\bibitem{CGHJK1996} Corless R M, Gonnet G H, Hare D E G, Jeffrey D J and Knuth D E, 1996 {\it Adv. Comput. Math.}{\bf 5}, 329 (1996); Corless R M, Jeffrey D J, and D. E. Knuth D E, 1997  in {\it Proceedings of the International Symposium on Symbolic and Algebraic Computation}
\bibitem{note0} The present calculations show that there is not a unique metastable configuration in the thermodynamic limit, contrary to what is suggested by Liu Y and Dahmen K, {\it Preprint} cond-mat/0609609. Of course, this has no incidence on the macroscopic properties discussed by these authors.
\bibitem{note1} In the nonequilibrium evolution, however, the one-spin-flip dynamics imposes a well-defined path among these states that depends on the field history. 
\bibitem{NS1999} Newman C M and Stein D L,  1999 {\it Phys. Rev. E} {\bf 60}, 5244
\bibitem{note2} For a related problem, see e.g. Mezard M and Parisi G, 2003 {\it J. Stat. Phys.}{\bf 111}, 1
\bibitem{PRT2008b} Rosinberg M L, Tarjus G and  P\'erez-Reche F J, in preparation
\bibitem{DRT2005} Detcheverry F, Rosinberg M L and Tarjus G, 2005 {\it Eur. Phys. J. B} {\bf 44}, 327
\bibitem{PRT2008a} P\'erez-Reche F J, Rosinberg M L  and Tarjus G, 2008 {\it Phys. Rev. B} {\bf 77}, 064422
\bibitem{B1998} See e.g. Bertotti G, 1998 {\it Hysteresis in Magnetism}, Academic Press, and references therein.
\bibitem{BMPREV2007} Bonnot E, Romero R, Illa X, Ma\~nosa L,  Planes A and Vives E,  2007 {\it  Phys. Rev. B} {\bf  76}, 064105
\bibitem{IRV2006} Illa X, Rosinberg M L and Vives E, 2006 {\it Phys. Rev. B} {\bf  74}, 224403; Illa X, Rosinberg M L, Shukla P and Vives E, 2006 {\it Phys. Rev. B} {\bf  74}, 224404

\end{thebibliography}
\end{document}